# Impossible?  Publication Quality Research with the Weakest 10% of Incoming Freshmen

Michael Courtney, Ph.D.,[1] and Amy Courtney, Ph.D.[2]

[1] U.S. Air Force Academy, 2354 Fairchild Drive, USAF Academy, CO, 80840

Michael_Courtney@alum.mit.edu

[2] BTG Research, PO Box 62541 Colorado Springs, CO 80920

**ABSTRACT**
Undergraduate research is widely regarded as a high impact practice.  However, usually only the highest achieving students are rewarded with undergraduate research opportunities.  This paper reports on the successful implementation of a student research program offering the weakest 10% of incoming freshmen opportunities to conduct original research in one of several science or engineering disciplines with the possibility of publication if the research and report meet a suitable standard, defined as earning an A on the final research project report in the introductory math course.  The opportunity has been offered now for two years to incoming cadets at the United States Air Force Academy who are placed in Basic Math.  The cadets placed in this course score in the bottom 5% of incoming cadets on the math placement exam.  During the second semester of their freshman year, cadets enrolled in Calculus 1 are also offered a similar research opportunity.  About 10% of cadets are enrolled in this course each Spring, the 5% who began in Basic Math and matriculate to Calculus 1 and the 5% who failed Calculus 1 in their first attempt.  During the first four semesters, the program has yielded 22 cadet papers which have been published or are currently under review and expected to be published.  This represents approximately 38% of the projects in the program, because the majority of the projects do not earn As and are not suitable for publication.  Over 80% of the cadet co-authors on the publication quality papers are minorities, women, and/or intercollegiate athletes.

"If you don't have the *ganas*, I will give it to you, because I'm an expert." - Jaime Escalante, Stand and Deliver

**Introduction and Background**
All entering cadets at the United States Air Force Academy are given a math placement test, which along with high school records and standardized test scores, are used to place each cadet in an appropriate math course.  The overwhelming majority of cadets (~95% most years) are placed into some level of Calculus.  A small minority (~5%) are placed in Basic Math to strengthen the math skills needed to succeed in the Academy's challenging technical core (including two semesters each of Physics, Chemistry, and Calculus and four semesters of Engineering courses).  Each class of incoming cadets is between 1000 and 1200 students, and the cohort placed in Basic Math is typically between 50 and 60 cadets, who are divided into three course sections.

The idea of a publishable research for  weakest 5% of incoming freshmen initially got a cold reception from some colleagues in the science, technology, engineering, and mathematics (STEM) fields.[1]  STEM outreach to the community is a popular activity at the Air Force Academy, including several programs and efforts aimed at underprivileged and underachieving audiences.  However, no one was initially eager to provide laboratory space or funding for a research program providing opportunities to the Air Force Academy's weakest students in math.  In spite of challenges, the number of publications and accomplishments by participants suggest we are succeeding in the mission of motivating and enhancing the quantitative skills of cadets with soft STEM backgrounds.

In spite of some negative reception, the authors and faculty mentors were motivated to persist by three main factors.  First, the Dean of the Faculty kept emphasizing that the purpose of research at the Air Force Academy is the education of cadets.  As a faculty member, MC is an enthusiastic and prolific research scientist, and it did not seem consistent with the mission of the institution to exclude cadets from his research.  However, as a

---

1  Some resistance was simply based on practical concerns such as "How are you going to do publication quality work with students still lacking in their Algebra and Trigonometry skills?" and "How are you going to fund the required laboratory space and equipment needs?"  However, we have also encountered something of an "old guard" mentality that seems to prefer a "sink or swim" mentality rather than reaching out to weaker students and providing greater opportunities to "swim."

Mathematics faculty member working in the Quantitative Reasoning Center under Student Academic Services, MC did not have easy access to the science and engineering majors who provide most of cadet research involvement, and he realized the need to carve out a new niche for cadet research participation. Second, some years before, AC had given a talk at West Point along the lines of "Everything I needed to know, I learned in Physics 201 and Physics 202" which pointed out how much quality science we had done over the years applying introductory physics and relatively simple mathematical analysis to a broad array of problems in ballistics, blast, and traumatic brain injury (Courtney and Courtney, 2007). We believed that interesting and relevant work could be performed with introductory level skills, and that the more interesting we could make the work for cadets, the less they would notice how hard they were working in growing their skills to meet the needs of the projects. Third, being people of faith, we believed that "Many who are first will be last, and many who are last will be first." (Matthew 19:30, New International Version) In the context of science and engineering, we had often seen much money and mental effort fail to solve a problem and an elegant and simple solution arise from an approach less blinded by complex computational details yet guided by the notion that the correct solution to many challenging problems is bounded by the solutions of two much simpler problems.

The authors' research backgrounds are in biomechanics, atomic physics, blast and ballistics, and our teaching backgrounds include a mix of both high school and college level physics, math, and biology. Having trouble selling the idea of research for the weakest 5% of students, one of the authors, (MC) applied for an internal Air Force Academy grant for a high speed (20,000 frames per second) video camera. The grant application described the projects the camera would be used for and also described his previous experience in the fields of blast and ballistics. The application also promised to use the camera for creating cadet research opportunities, but after a few false starts, the application did not mention the cohort of cadets who would participate in the research. Together with some brainstormed ideas for inexpensive experiments using equipment borrowed from consulting relationships, analysis and interpretation of some publicly available data, and ideas for analyzing and publishing some data sets that the authors happened to have on hand, and we managed to put together a list of exciting project opportunities for the cadets in Basic Math in fall 2010. Once MC got word to his chain of command (USAFA Student Academic Services) that 7 projects had earned As the first semester and were certain to be published, Student Academic Services was able to provide an additional $12,000 for the purchase of materials and equipment. When federal budget cuts threatened the research program, MC appealed to his wife AC (majority owner of BTG Research) who approved the necessary cooperative agreement to continue.

However, since no one is lining up to throw megabucks at research being conducted by students with soft STEM backgrounds, economy is essential to the program. Each cadet research project averages in the neighborhood of $800 in direct costs, and since only about 1/3 of the projects are ever published, the effective average costs of each publication is near $2400. Most STEM faculty would have difficulty doing much with direct costs of $2400 per publication, much less figuring out how to have the weakest students participating in the research.

The authors were also motivated to offer research opportunities, because we recall the delightful eye opening of our own early research experiences. Basic Math and Calculus 1 with the "trailers and failers" (as the Spring Calculus 1 cohort is sometimes called) have not historically been motivating experiences for cadets, nor are these courses known for being well populated with cadets with high levels of motivation for the skill based drills that turned them off to mathematics in high school. The strategy was simple: maybe a research opportunity with an explicit promise of publication could find the "on button" for some students where other approaches had failed. Incorporation of research into Basic Math also coincided with an adjustment of the focus of the course away from "pre-Calculus" to being a preparation for the USAFA technical core as a whole, and an emphasis on mathematical modeling rather than the more traditional course focused on mechanical manipulations of equations and expressions. The focus was shifted from repetitive mathematical drills to word problems emphasizing that the laws of nature are written in the language of mathematics, and that mastery of mathematics is essential to the mastery of science and engineering required of officers in the modern Air Force.

From the beginning it was clear that attempting for every project to yield a a publication in a peer-reviewed journal would likely extend the time line for the positive feedback too long and also prohibit us from offering a project selection closely aligned with cadet interests. The Air Force Academy has built in mechanisms where every publication is required to receive feedback and consideration through both faculty and administrative review and approval. Therefore, when brainstorming for possible project ideas, rather than worship at the altar of peer review, we consider the uniqueness of the idea, the scientific and mathematical learning objectives, the expected level of cadet interest, the potential to be completed in a single semester, and the potential applicability of the project to Department of Defense interests.

Many outstanding cadet reports have appeared as magazine articles in Precision Shooting, Varmint Hunter Magazine, and Rockets (The Official Magazine of the Tripoli Rocketry Association). Other papers are published as Department of Defense Research Reports or Technical Reports through the Defense Technical Information Center. Some project papers are published via the online repositories of the Cornell University Library (arXiv) or the Nature Publishing Group (Nature Precedings). A minority of project papers are submitted either to mainstream peer-reviewed journals or undergraduate research journals.

Some preliminary comments on possible publication venues are made in the main project document describing the potential project offerings, and there are further discussions between faculty and cadets as the project and report progress. Once they understand the trade-offs, cadets themselves choose the venue for publication submission. Some cadet groups want to have a glossy magazine to show off to family, friends, and colleagues. Other cadet groups prefer the fast appearance of published works in the internet repositories and the ability to send a link to parents and friends or post a link to their publication on FaceBook. Many cadets want the permanence and reputation garnered by publishing as a Department of Defense Research Report. A few (especially aspiring STEM majors) come to appreciate the distinction between the "grey literature" and peer-reviewed journals and are willing to wait the extra time required to guide a paper through the process. Some groups initially aspire to make a paper of marginal scope or impact into a peer-reviewed paper and then settle for a choice in the grey literature after a rejection or two. Occasionally, a group will produce two project reports, an informal report intended for a hobbyist magazine usually using English units and lacking in technical detail, and a more formal report suitable for a science or engineering journal.

**Program Implementation**
The opportunity to participate in the program is offered to cadets enrolled in Basic Math in the fall of their entering year and to cadets enrolled in Calculus 1 in the following spring semester. The cohort of cadets in spring Calculus 1 is dominated by cadets matriculating from Basic Math in the fall and cadets who failed Calculus 1 in the fall. There are about 120 cadets in a typical spring offering of Calculus 1, making up 10-12% of the freshman class. Cadets are in peril of disenrollment if they have not passed Calculus 1 by the beginning of their Sophomore year because of the way pre-requisites line up for the sequence of courses in the required technical core. There is a three week summer offering of Calculus 1 for cadets who have not yet passed the course, but the downstream success of cadets who pass Calculus 1 in the summer is abysmal (< 50%).

Instructors announce the research and publication opportunity when discussing course content on the first day of class. Cadets are guided to a main project document which describes the project opportunities and administrative details. It is emphasized that only A quality work will be approved for publication, and that a consistent effort over time is key to success. The main project categories are discussed, along with reference to past project successes and a few of the more exciting current opportunities. Students are told that they should begin considering topics and possible project partners as these choices need to be made by the 9th or 10th lesson. The room usually buzzes with excitement, and there are a lot of questions split between clarifying administrative details and asking for further information on specific topics.

Once they are aware of the possibilities, groups sometimes are eager to discuss related project topic ideas. The faculty mentors try and facilitate ideas while being mindful of resource constraints, good scientific design, and the required analysis skills. Before receiving input from faculty, cadets are often inclined to come up with project ideas with too many independent variables and overly broad and ambitious to be accomplished in a single semester. Invariably, cadet ideas include an exciting experiment but underestimate the subsequent analysis and composition efforts to produce an excellent research report. Since the project is coincident with a mathematics course, we offer a good mix of projects only requiring mathematical computations and data analysis and projects with a required experimental component that will ultimately be much more time consuming. Some semesters, all cadets have been required to do a custom project with the possibility of publication for projects earning As. Other semesters, we offer a default project with course related learning objectives for cadets who prefer not to pursue a publishable project. We have come to prefer this approach, as it has the result of requiring fewer resources to support projects where the cadets are not motivated to become published authors.

**Rolling Up the Sleeves**
A high level of initial inspiration and motivation are not always sufficient to ensure the level of effort and quality work over the course of the project that is required to yield excellent results, and this cohort has a tendency to underestimate the level of effort required.  Cadets are encouraged to work closely with their faculty mentors who take care to make themselves available and accessible to cadets, but cadets have the responsibility to initiate requests for faculty help and to meet project milestones.  Groups are encouraged to reach out to a faculty mentor soon after selecting a project topic to lay out a concrete plan for completing the experiment (if necessary), acquiring the data from a public or private source (if necessary), analyzing the data, and drafting the report on a schedule to provide the required course deliverables in a timely manner.  Project groups which are inclined to procrastinate are allowed to, but are gently encouraged to get working.

The first course deliverable is project Part 1, which should be a complete analysis of the data including graphs, tables, and all the essential mathematical computations.  Captions are required for the figures and tables, but it is not required for the cadets to have drafted any of the text.  Usually there is a lot of project activity the week before Part 1 is due, with several groups waiting until the night before.  Groups who were working with a faculty mentor and getting regular feedback on their work have usually produced correct and complete analysis, and the only flaws in their Part 1 work are typically small font sizes in their figures once imported into a document and lack of uniform formatting or perhaps inconsistent labeling on figures and tables.  Many project groups who wait until the night or two before Part 1 is due often have incorrect or incomplete analysis, and the figures and tables are poor if they are included at all.  Over time, we have increased the point allotment of Part 1 up to 30% of the total project points in order to encourage better work at this stage, because a good start is key to quality work going forward.  Occasionally, data collection may be continuing when the Part 1 deliverable is due.  Rather than extend the deadline which might encourage procrastination, Part 1 is accepted and graded based on whatever data is already in hand at the time.

Project Part 2 is expected to be a complete intermediate draft, including abstract, introduction, method, results, discussion, and bibliography.  Most projects stick closely to this basic outline, though working with a faculty mentor there are occasional differences in the basic outline.  For example, some fisheries science papers are organized as introduction, fish 1, fish 2, . . . , discussion, etc.  Here, a big issue is that this cohort of cadets not only has significant weaknesses in mathematics, they manage to enter college without well formed writing skills either.  Groups who work closely with faculty grow in their skills and produce satisfactory work.  Groups who are winging it the night before often produce abysmal work.  Students in this group often do not bother to compare the analysis they are resubmitting on Project Part 2 with the feedback received in Part 1, and it is common that both computational and formatting errors marked in Part 1 are repeated in Part 2.  We've settled on 20% of the project points for Part 2.

Part 3 of the project is cadets' last chance to get it right and produce a publishable work product.  It is the final report for a grade.  Most groups who make a consistent level of effort over the semester earn As, because they seek and are sensitive to feedback from faculty mentors.  Some of the more talented groups whose approach was more of a week long effort right before turning in each deliverable manage to eke out As.  Most of the groups whose project was pursued in fits and starts earn Bs, Cs, and Ds.  There is the occasional group who shows tremendous effort over the course of the semester, yet still only manages B work on the final report due to pre-existing weaknesses in mathematical, organizational, or writing skills that could not quite be completely overcome in the available time.  In these cases, cadets who persist and select a second research opportunity the following semester have always earned As and produced publishable work on the second try.

**Group Dynamics**
There have been several occasions where one project partner was highly motivated and worked hard, but other students in the group were contributing little.  The first choice of faculty mentors is to gently encourage cadets to work these things out among themselves.  Faculty mentors have seen numerous occasions where peer pressure is more motivational than faculty guidance.  One occasion in particular stands out where a motivated group member picked up the graded draft for Part 2 walked over to the project partner and slammed it down on the desk in front of him.  Clearly, this partner had let the group down in some way.  We suspect that the conversation was continued in private.  We were not privy to the detailed group dynamics, but we did see a rapid improvement in both that cadet's classroom work, as well as a much improved work product on project Part 3, which with additional editing was eventually published in a peer-reviewed journal.

On occasion, remediation efforts are unsuccessful, and a motivated group member may want to split from less motivated members. Faculty mentors allow group splits in these cases, and in every case, a project paper of excellent quality has resulted. In one case, both remaining groups were assigned the same project, and the motivated member turned in excellent work, while the unmotivated member independently produced poor work. In a second case, the group split provided sufficient motivation for the originally unmotivated member to improve the quality of his work, once he realized that he was not going to get a good grade or a publication merely riding the coattails of his partner. Since this was a fish project which was divided by species, the two A level papers were relatively easily edited into a single paper and a quality publication resulted with both students listed as co-authors.

There have been several occasions where a symbiosis of academic strengths, personality, and motivation produced both excellent work and growth, and where groups continued to grow over the course of two projects completed over their freshman year. Once a group gets into a positive motivational feedback cycle, the progress is truly satisfying to watch, and the role of the faculty mentor fades to providing technical advice as the group charges toward excellence. In one case, the group added a third member for their spring semester project, and the additional technical skills of the new member allowed a truly impressive volume of quantitative analysis to be completed. The organizational and writing skills had been honed sufficiently by the two original group members in the first semester that the final paper was excellent and only in need of minimal additional editing before publication. In a second case, a group of two highly motivated cadets added a third group member for their second semester project. Early in the second semester, it appeared as if the positive influence of the two original members would successfully mentor the new member toward improved in work habits, but as the grind of the semester wore on, the new member fell back into old habits and attempted to take credit for his partners' work. Faculty mentors take care to ensure that each listed co-author has made significant contributions to the final work product, and in cases such as this one, group members who have not made substantial contributions are excluded from authorship on the published paper.

We've tried to recognize the recipe for group members having great positive impacts on each other, but we have not yet identified any keys to reliably successful interpersonal dynamics. The highly successful groups have been the same gender and mixed gender, same sports team and different sports teams, intercollegiate athlete and non-intercollegiate athlete, mixed race and same race, different cadet squads and the same cadet squad. We have noticed that groups with a cadet who was an enlisted airman before enrolling at the Air Force Academy do tend to succeed more often than groups where all the cadets are straight out of high school. Likewise, groups with hockey players, who also tend to be older due to several years playing minor league hockey, also tend to perform well. Older students who have experienced several years of life outside of school seem more likely to fully recognize the great opportunities of being a cadet at the Air Force Academy and in conducting original research as a freshman, and their maturity often empowers their leadership abilities in imparting their enthusiasm to other group members. We believe these peer interactions have been a more powerful motivator than interactions with faculty mentors.

The most negative outcomes occur when cadets procrastinate and then attempt to piece together a project deliverable by assigning sub-tasks to be pasted together by a member acting as an editor at the last minute. The project document and faculty mentors discourage this "divide and conquer" approach, but it seems to be an ingrained feature of human behavior when under time pressure and competing demands. We try and teach that good science can be conducted with a division of labor, but the best work has all group members having ample opportunity to review and offer constructive feedback on every area of the project: every graph, every bit of analysis, every paragraph of text, and every use of references. In some project groups, honor cases have arisen when desperate group members cut and paste text from published sources without proper citations.

In addition to forwarding potential plagiarism to the cadet honor system, the Math Department has given academic penalties for academic dishonesty, and the program leadership has immediately disqualified projects from possible publication if there is substantial evidence of academic dishonesty in work submitted for a grade. Some faculty have considered this too harsh and recommended that such situations represent an opportunity for growth where the seriousness of cheating can be effectively communicated with the possibility for publication preserved. However, the program leadership feels that the responsibility for dishonesty must be commensurate with the value of the opportunity, and that termination of the opportunity to publish is needed to effectively communicate the seriousness of the offense and maintain the integrity of the program (See for example, Williams and Courtney, 2011).

**Topic Area: Ballistics**
Ballistics is a popular topic area. Many cadets with interests in the profession of arms are enthusiastic about the study of ballistics, and there are many topic possibilities that are accessible with first year college math and science skills. Since the Basic Math course emphasizes mathematical modeling, projects with applications of mathematical modeling are emphasized. These projects also have ample opportunity to master the concept of functional composition, because there are many cases where analysis requires the output of one function to serve as the input to another. Most ballistics projects include enough rudimentary physics and engineering that cadets are simultaneously getting an opportunity to grow in these skills in improved preparation for the technical core.

Projects include offerings in internal ballistics where cadets have conducted intensive studies of barrel friction and rifle primer performance. Studies in barrel friction are appealing because they have immediate and obvious military applications. Yet the basic technique was stumbled upon by accident. MC had collected a large data set for a study in bullet stability (Courtney and Miller, 2012a and 2012b). We had given consideration on how to involve cadets in the bullet stability studies, but decided that the math was beyond first year capabilities and had seen an initial attempt at cadet involvement fail the previous year. However, MC's preliminary analysis of the data set in terms of muzzle energy vs. powder charge revealed extremely linear high correlation coefficients ($r > 0.9999$) and a clear interpretation of the vertical intercept as the energy lost to friction. The cadet group that picked the project with the task of completing the analysis produced a great paper (Boyle et al. 2012a) demonstrating the utility of the method, and the paper was also published in Precision Shooting magazine (in print, March 2012) and Long Range Hunting (an on-line magazine, March 2012). The group followed-up with a study of the effectiveness of bullet coatings as purported lubricants in the spring semester (Boyle et al. 2012b), and other projects are moving forward at the present time studying the friction of different military bullet designs, and the impact on barrel friction of lead free rifle primers.

The idea to study rifle primer performance with high speed pressure transducers was also stumbled upon by accident. A defense contractor had stopped by the authors' laboratory for assistance validating a blast wave measurement system. During the course of the day, we needed a simple and easy method for producing blast waves. MC walked over to the reloading bench and put a few primers in empty brass cartridge cases. We attached a pressure transducer to the end of the rifle barrel, loaded the primed (but otherwise empty) cartridge case, and pulled the trigger. The pressure transducer measured a near perfect Friedlander waveform at the end of the rifle barrel. In that unexpected moment, we realized we had invented both a table top shock tube that would prove useful in applying a realistic blast wave profile to small samples (Courtney and Courtney, 2010; Courtney et al. 2011) as well as a method for quantifying the power and consistency of rifle primers without the confounding effects of alternate methods (Courtney and Courtney, 2011; Courtney et al. 2012). This tool has also been used in several cadet projects studying blast wave transmission through materials that were important learning experiences for cadets and the faculty mentors, but these projects did not yield publications.

Studies in external ballistics have included extensive work in aerodynamic drag, both working with different drag models and in analyzing data relating to whether products meet their published drag specifications. One impressive project was testing the hypothesis that polishing a rifle bore would reduce the aerodynamic drag on bullets fired from that rifle (Bohnenkamp et al. 2012a). A second project compared manufacturer specifications for aerodynamic drag with independent measurements (Bohnenkamp et al. 2012b). A third project presented analysis of original aerodynamic drag measurements (Halloran et al. 2012). Most of the work in ballistics is facilitated by a ballistic calculator that performs complex numerical integrations to compute trajectories from given inputs or computes ballistic coefficients from given measurements. The computational burden of these projects is consequently fairly low, but the requirements for conceptual mastery of functional inputs and outputs, as well as for understanding inverse functions and functional composition is high.

Studies in terminal ballistics range from terminal performance and bullet effectiveness to studying armor effectiveness and bullet penetration. For example, one project studied whether or not bullet penetration in armor testing would depend only on the material and its thickness, or if it also depended on the total mass of the sample (Haight et al. 2012). Cadets reported the result that testing on armor samples smaller than the actual fielded armor yields overly optimistic estimates of the impact velocities the armor is able to stop. Execution of the project required fitting experimental data to logistic type curves to determine the V50s, as well as simple statistics, and physical reasoning to explain the results.

**Topic Area: Blast**
Cadets are initially intrigued by projects in the field of blast, but many shy away due to the perception that the mathematics of blast related projects will somehow prove prohibitively difficult. We've designed the blast projects to be accessible with the skills taught in Basic Math and Calculus 1, but the ability of cadets to visualize the physical scenarios in ballistics and rocketry, but not in blast have made blast a relatively unpopular topic area. The most successful blast projects were conducted by cadets whose enthusiasm and zeal overcame their apprehensions. This is also the field that has yielded the most peer-reviewed cadet publications.

The groups completing blast projects simply analyzing and presenting test data acquired by others have never produced publication quality work. The analysis requires significant attention to detail and is often regarded as tedious with several conceptual difficulties that seem particularly challenging to those who did not participate in the actual experiment. The physical meaning of concepts like sensor calibration factors, high-speed digitizers, time constants, and transmission percentages seem to get lost in the task management while struggling with other issues like labeling axes in graphs and describing an experimental method that was not actually witnessed.

In contrast, the groups completing blast projects who actually participated in the experiment have never failed to produce publication quality work, even in cases where the volume of analysis effort greatly exceeded the average for course projects. In one project, cadets used high-speed video to determine the deflagration velocity in a mixture of oxygen and acetylene (Armacost et al., 2011). Since a similar demonstration is often presented as an explosion in chemistry classes, the distinction of the event as a fast deflagration rather than a detonation was published in The Chemical Educator. Of course, follow-up studies are available in both other gas mixtures and in confined mixtures, as deflagration velocities and deflagration to detonation transitions have several important possibilities in physical chemistry and military applications. Another great cadet project determined that prior blast exposures seem to increase the blast wave transmission of cranial bone in later exposures (Courtney et al., 2012b). A third cadet project used high speed video to quantify the magnitude and timing of momentum transfer to spheres from oxy-acetylene based shock tubes (Her et al., 2012). This experimental result confirmed earlier theoretical predictions (Courtney et al., 2012a) that momentum transfer is dominated by the blast wave itself and that the effect of venting gases (known as the "jet effect") is negligible in this shock tube design.

**Topic Area: Rocketry**
Rocketry is a popular topic area. Several project teams have select a rocketry topic in earnest, but to date there has only been one published paper (Hedrick et al., 2011). The quantitative analysis of rocketry projects usually centers around computation of the physical quantities of impulse and specific impulse from force plate measurements of a specific rocket design and fuel. These computations are technically related to Calculus 2, so they are advanced relative to the skills mastered by the students. However, the software tools that are recommended (a spreadsheet, Graph.exe, and the force plate operating software) all automate the computation with no real mastery of Calculus 2 being required, other than the idea that the impulse is the area under the curve of force vs. time. Faculty mentors are willing to sit down with rocketry project teams and walk them through the computations for the first few rocket motor firings. However, what we have seen is that while cadets enjoy the rush and thrill of the actual testing of rocket motors, many have lacked the work habits needed to complete the analysis in a timely manner. They seem to underestimate the skills and time required to complete a proper quantitative analysis and quality report and begin each part of the project too late to meet deliverable dates.

**Topic Area: Product Quality Testing**
The idea of catching a product manufacturer in an exaggerated product specification is appealing to cadets. As discussed above, several successful ballistics projects, and the only published rocketry project to date have had the central project focus comparing measured values with a published specifications. We also had a published project where the cadets measured the breaking strength of fishing line with a force plate and compared different brands with their published specifications (Haight et al., 2012b). As future officers in the Air Force, many cadets will have roles in acquisition and product testing and selection, so these projects reinforce the need to keep vendors honest about specifications, as well as the skills to design experiments to measure specifications and the quantitative skills needed both to interpret specifications and test results in light of experimental uncertainties to determine with some specified confidence level whether or not the product is within specification.

**Topic Area: Quantitative Fisheries Science**
When designing the QRC Cadet Research Program, we wanted include ample topic choices that might have more appeal to cadets less interested in physical science and perhaps not yet strongly motivated by projects with more obvious applications in the profession of arms. Yet, an area was needed in which a faculty member either already had or could quickly develop the knowledge base to effectively serve as a faculty mentor and where the quantitative analysis was sufficiently challenging, but not prohibitively difficult for the cohort of students being served. The authors had been working with their children for several years weighing and measuring fish for school science projects, and in the process of double checking a child's work, AC had discovered obvious errors in some weight-length parameters at Fishbase.org, an online database of quantitative data in fisheries science. A lot of quantitative fisheries science involves power law models, linear models, and ratios, so the math is amenable and useful for illustrating important concepts and skills in Basic Math. It was a natural fit.

During the first semester of the program (fall 2010), the majority of the projects earning As (four out of seven) were in the area of quantitative fisheries science. One project was based on a data set on the weight and length of cutbow trout acquired by the authors' family in the summer and fall of 2010 (Parker et al., 2011). A second project on channel catfish was based on a data set acquired by the authors' family in the spring and summer of 2009 (Keenan et al., 2011). A third project on several freshwater species was based on a data set acquired by the authors' family in 2008 (Dexter et al., 2008). The fourth published project (Cole-Fletcher et al., 2011) in fisheries science explored and documented the errors in weight-length parameters in FishBase.org that had originally been discovered by AC. The editors of FishBase.org have downplayed the impact of this project report in their reviews and online discussion forums, but they have instituted systematic efforts to double check the weight-length parameters of freshwater species and many errors have been fixed. We had notified the editors of the errors cadets discovered months in advance of publishing the paper, but the FishBase did not begin fixing the errors (at least the fixes did not become public) until after the paper was published.

Data sets for analysis-based projects are often recycled and offered as projects in following semesters in cases where the project group fails to earn an A, but the high success rate for fisheries projects created something of a bind, because the program had exhausted most of the original data sets we had on hand, and data sets with sufficient sample sizes are labor-intensive to acquire. Fortunately, the study documenting the errors in the weight-length parameters at FishBase.org noticed that the majority of errors in that database had a single source in common (Carlander, 1969), which is a freshwater fisheries handbook that has been cited over 1500 times and is widely depended upon. This gave us the idea that future projects could be of service to the fisheries science community by documenting and (where possible) correcting the errors in that widely cited source. A number of cadet projects have made attempts at this. Unfortunately, success in these projects requires more creative thinking and attention to detail than many cadet groups are willing to expend. The mistakes are easy enough to find by plotting the purported weight-length formulas for a small group of related species, but fixing the mistakes requires assembling data that is scattered over many different tables in the entries for a given species, and to date only one project group has demonstrated the required level of creativity and attention to detail to yield a publication quality project (Daviscourt et al., 2011).

Some of the best projects in fisheries science have been submitted to peer-reviewed journals, but the standard for publication in most fisheries journals is a high level of relevance rather than merely sound science and good quantitative analysis. Most "survey" type of studies and follow-up analysis on prior work are relegated to the "grey literature." Consequently, all the cadet projects were eventually published as reports in the grey literature rather than in peer-reviewed fisheries journals.

The success of the fisheries science projects in the first two semesters inspired considerable effort during 2011 to acquire new original data for additional fisheries science projects in the 2011-2012 school year. Once again, multiple projects earned As and were submitted for publication. Having not previously had cadets submit their work to undergraduate research journals, several groups who earned As were encouraged to submit to these journals to test these waters as possible publication venues. Unfortunately, through the process we have learned that many of these journals are in various states of editorial disarray, and many months after initial submission, none of these journals has been able to provide even a single peer-review of the project papers that have been submitted. Consequently, we expect that we will soon be withdrawing these papers and publishing them in the grey literature. It was not an edifying or encouraging process for students to submit their hard work and hear nothing in reply for many months. Interested parties should contact MC via email for more specific information regarding unresponsive undergraduate research journals.

The time and cost required to obtain original data sets and the required attention to detail to produce publishable analysis of the Carlander (1969) handbook, started us thinking about alternate approaches to obtain data for student research projects. In 2010, we had written to one state wildlife agency requesting some of their fisheries survey data, and after some initial positive indications, they later sent us a lawyerly letter denying our request. However, because we often spend hours each week doing background literature searches to develop new ideas, we eventually stumbled upon a grant award for a state wildlife agency to develop a database for the specific purpose of facilitating data sharing with other government agencies and research institutions. This state had completed development of the system, and quickly responded to our written request with three spreadsheets with thousands of data points relating to three of the most important fisheries in that state. Fall 2012 is the first semester we're offering projects analyzing data from the state agency, but we expect positive results. We hope this approach can significantly reduce the time and costs of conducting our own fisheries surveys to acquire original data for cadets to analyze.

**Discussion**

There are ample opportunities for presenting undergraduate research, so one may wonder why we have geared the program toward publication instead. First of all, it is important that the entire project commitment be confined to a single semester. Cadets in Basic Math and spring Calculus 1 have tremendous time constraints due to the volume of coursework and additional training requirements. Guiding a paper through the publication process after the semester in which the project is completed is much less time consuming and has more flexibility regarding when that time is invested. Second, there simply are not the funds available to provide for conference presentations for a significant fraction of projects that earn As. Third, while presenting at a conference is a valuable experience for students, often the results of the work are lost and not available in the archival literature. Our publication model allows future research groups, including student groups, to reinterpret and build on past project results. Fourth, many students in this cohort are more motivated by the opportunity to publish than the opportunity to present. Finally, we were slightly concerned that elitism expressed at undergraduate presentation venues dominated by the "best and brightest" upper class STEM majors might not be motivational for the cohort from Basic Math. Publishing provides ample positive feedback, and cadets are often lauded by their cadet squadrons, sports teams, and departments when an announcement of their published work appears in the Dean's Weekly newsletter.

One may also notice that faculty mentors appear in the list of authors on most cadet publications. In the cases where cadets performed the data analysis and drafted the paper, it is necessary for at least one of the listed co-authors to take actual responsibility for having acquired the data. In addition, even in cases where cadets performed the bulk of the data collection, it is usually the case that the level of contribution of the faculty mentor(s) to the original experimental design, guiding the data analysis, and polishing the final manuscript warrants inclusion as co-authors. In some cases, as the paper moves through the peer-review and revision process, the contributions of more senior collaborators increase to the point of warranting first authorship, though this is uncommon.

We wish we could provide better guidance on how similar programs might be successfully implemented at other institutions, but at the heart of the matter, we believe that this may be an educational challenge for which there may be no generally applicable recipe given the variation of academic and research cultures across institutions and the variation of student cultures and needs. Obviously, one needs faculty with sufficient breadth of knowledge to mentor the research, one needs an institutional culture willing to reward faculty for such a non-traditional approach (or faculty who can be content if the required efforts negatively impact their retention, promotion, or tenure), and one needs adequate funding and facilities. But this kind of checklist misses the point that imparting a love for science, engineering, and mathematics inherently requires a deep love for these subjects and a deep desire to share these passions with the next generation.

Several facets of the program can rob motivation of faculty mentors if one is careless. Only 38% of cadet projects earn As and are suitable for publication. However, it would be careless to view papers earning Bs and Cs as failures. These projects serve to impart many important learning objectives in real world science and engineering problems. Furthermore, these projects assist students in growing in their expectations regarding the level of effort required to produce excellent work. Students who have been disappointed with Bs in their Basic Math project have raised their level of effort and gone on to earn As in their projects the second semester. The expense and effort of producing data sets and project ideas seems wasted on students who squander the opportunity and do not even manage C work, or worse yet, cheat in their attempts to get a good grade. It is not clear to us how students can arrive at the conclusion that we want them to succeed so badly that we are willing to give away grades or close our eyes to academic dishonesty. We suspect that earlier teachers may have done these students the disservice of gifting grades and winking at cheating. After all, the weakest 5% of students at an institution with high admissions standards probably did not obtain that distinction by being given commensurate grades every time they failed to demonstrate mastery of the required material.

Every instructor of students earning deficient grades is required to write a comment card on deficient students and must recommend to the Air Force Academy whether or not to retain deficient students. Recognizing the required taxpayer investment of $100,000 per year and the improbability of students who remain deficient in math skills succeeding in the challenging technical core, MC may well recommend more students not be retained than any other USAFA faculty member. It is hard to write comment cards recommending disenrollment, and doing one's duty here rarely wins any friends, especially when it comes to student evaluations. Rather than pretend that cadets have a chance to succeed after their window of opportunity is closed, we look for ways to inspire and motivate students while the window of opportunity is still open.

The authors and other faculty mentors in the program believe that this cohort of students will respond better to a firm hand and high expectations to help them move beyond common freshman misconceptions (see Courtney et al., 2007) rather than catering to student desires which can lead them to conclude that the actual standards for success are lower than advertised (see Courtney and Courtney, 2012c). Lowering standards only communicates disbelief in student abilities to succeed. Students who rise to the challenge of meeting high standards grow not only in their quantitative problem solving abilities, but also in work habits, self-esteem, and confidence that are essential to succeeding in downstream challenges. The program director (MC) who grades most of the projects is strongly disposed to assign projects a grade of B if they are on the borderline between an A and a B, though he consults with other faculty and he is willing to assign As when other faculty determine that projects have met an expected level of excellence and require only a little polishing to produce publishable work.

The attitude of the program director (MC) is often similar to Jaime Escalante as portrayed by Edward James Olmos in "Stand and Deliver." I have tremendous expectations regarding the ability of this cohort of students to succeed in any academic challenge they face. At the same time, I realize that with a challenging technical core ahead of them, their window of opportunity to master basic Algebra and Trigonometry skills is narrow, as cadets are expected to earn over 140 credit hours in eight semesters. As you digest our report on the research program, we recommend watching a video of "Stand and Deliver." Consider the miracle of what Jaime Escalante accomplished with the impossible challenges he faced. Then find the *ganas*[2] to do impossible things with your students!

---

2  In the movie, the translation given for *ganas* is desire. "Sacrificial desire" is a more appropriate translation in the context. In the movie, both teacher and students demonstrate sacrificial desire and give up much time and competing interests to accomplish surprising things. In the QRC Cadet Research Program, faculty mentors often exceed a 40 hour work week in support of the program, work on projects in basement and garage workshops, and pay some expenses out of our own pockets. The sacrificial desire demonstrated by cadets is even greater. Social activities and internet distractions are curbed. Cadets give up rare free weekend days to conduct experiments when free time is unavailable during the week. In later semesters, cadets have also given up coveted spots on intercollegiate athletic teams to ensure the level of academic success required to meet their greater goal of serving as officers in the United States Air Force.

**Acknowledgements**
Internal support for the research program has been provided by the Institute for Information Technology Applications at the United States Air Force Academy, as well as by the Office of Academic Affairs and Academy Registrar. External support has been provided by BTG Research and Force Protection Industries. Dr. Beth Schaubroeck (USAFA/DFMS) provided tremendous encouragement and allowed the beginning of the program through inclusion in the Basic Math course project. Other USAFA/DFMS faculty have also assisted in their roles as cooperating course directors, instructors, and reviewers of project reports. Dr. Lubov Andrusiv, Mr. Dave Summer, and Mr. Tom Slusher (USAFA/DFR) have served as faculty mentors in the program, with Dr. Andrusiv and Mr. Summer also providing advice and review services for projects not being directly mentored. MC deeply appreciates the flexibility of his chain of command, Dr. Tom Mabry and Dr. Dean Wilson (USAFA/DFR) regarding the work hours and locations needed to maintain the program, as well as their willingness to allow MC to aggressively pursue the program with so little material support. MC also appreciates Dr. Dean Wilson and Dr. Tom Mabry for creating and maintaining such a fabulous work environment in Student Academic Services. This team of faculty and staff is unparalleled. We are also grateful to Dr. Julie Tetley (USAFA/DFR) for suggesting we write a paper to describe the program and for reviewing the manuscript. We also deeply and truly appreciate the Air Force Academy's commitment to academic rigor for all cadets, and the broad consensus that academic rigor is essential in producing officers of character. In particular, support for academic rigor from the Athletic Department (directed by Dr. Hans Mueh) is tremendous and unparalleled at Division I schools. Distribution A. Approved for public release. Distribution unlimited. The views expressed in this paper are those of the authors and do not necessarily represent those of the U.S. Air Force Academy, the U.S. Air Force, the Department of Defense, or the U.S. Government.

**References**
Michael A. Armacost, Ayesha Pladera, and Michael Courtney, "Measuring Deflagration Velocity in Oxy-Acetylene with High-Speed Video," The Chemical Educator, October 2011.

Emily Bohnenkamp, Maurice Motley, and Michael Courtney, "Does Polishing a Rifle Bore Reduce Bullet Drag?" DTIC, 2012a, also Precision Shooting, April 2011.

Emily Bohnenkamp, Bradford Hackert, Maurice Motley, and Michael Courtney, "Comparing Advertised Ballistic Coefficients with Independent Measurements," DTIC, 2012b, also Varmint Hunter Magazine, Winter 2012.

Patrick Boyle, Alexander Humphrey, Spencer Proctor, and Michael Courtney, "Measuring Barrel Friction in the 5.56mm NATO," DTIC, 2012a.

Patrick Boyle, Alexander Humphrey, and Michael Courtney, "Friction Effects of Common Bullet Coatings in 5.56mm NATO," Precision Shooting, August 2012b.

Kenneth Carlander, "Handbook of Freshwater Fishery Biology, Volume 1," The Iowa State University Press, Ames. Iowa. 1969.

Simeon Cole-Fletcher, Lucas Marin-Salcedo, Ajaya Rana, Michael Courtney, "Errors in Length-weight Parameters at FishBase.org," Ecology , Nature Precedings, April 2011.

Michael Courtney, Norm Althausen, and Amy Courtney, "Five Frequently Fatal Freshmen Physics Fantasies," Physics Education, January 2007.

Amy Courtney, Lubov Andrusiv, and Michael Courtney, "Oxy-acetylene Driven Laboratory Scale Shock Tubes for Studying Blast Wave Effects," Review of Scientific Instruments, April 2012a.

Amy Courtney, Alivia Berg, George Michalke, and Michael Courtney, "A History of Blast Exposure May Effect the Transmission Properties of Cranial Bone," Experimental Mechanics, July 2012b.

Amy Courtney and Michael Courtney, "Everything I really needed to know I learned in Physics 201 and 202," Invited Colloquium at the Department of Physics and Nuclear Engineering, United States Military Academy,

November 2007.

Michael Courtney and Amy Courtney, "A Table-top blast driven shock tube," Review of Scientific Instruments, December 2010.

Amy Courtney and Michael Courtney, "Comparing Blast Pressure Variations of Lead Styphnate Based and Diazodintrophenol Based Primers," Weapons Systems Technology Information Analysis Center Journal, October 2011.

Michael Courtney and Amy Courtney, "Science and Math Education: Who is the Customer?" Cornell University Library, August 2012c.

Amy Courtney, Alivia Berg, George Michalke, and Michael Courtney, "Development and Characterization of Laboratory Scale Shock Tubes for Studies of Blast Wave Effects," Second Research Symposium on Traumatic Brain Injury, May 2011.

Michael Courtney, Joshua Daviscourt, Jonathan Eng, and Amy Courtney, "High-speed Measurement of Firearm Primer Blast Waves," Cornell University Library, March 2012.

Michael Courtney and Don Miller, "A Stability Formula for Plastic-Tipped Bullets," Precision Shooting, January 2012a.

Michael Courtney and Don Miller, "A Stability Formula for Plastic-Tipped Bullets," Precision Shooting, February 2012b.

Mercedes Dexter, Kyle Van Alstine, Michael Courtney, and Ya'el Courtney, "Demonstrating an Improved Length-weight Model in Largemouth Bass, Chain Pickerel, Yellow Perch, Black Crappie, and Brown Bullhead in Stilwell Reservoir, West Point, New York," Quantitative Biology, Cornell University Library, July 2011

Christine Haight, Kadie McNamara, and Michael Courtney, "Does V50 Depend on Armor Mass?" DTIC, 2012a.

Christine Haight, Kadie McNamara, Kathleen McQueeney, and Ya'el Courtney, "Comparing Measured Fluorocarbon Leader Breaking Strength with Manufacturer Claims," Cornell University Library, February 2012b.

Alex Halloran, Colton Huntsman, Chad Demers, and Michael Courtney, "More Inaccurate Specifications of Ballistic Coefficients," DTIC 2012.

Garret Hedrick, Elliot Beski, Timothy Lopez, and Michael Courtney, "Testing Estes' Thrust Claims for the A10-PT Motor," Rockets, April 2011.

Hunter Her, Amy Courtney, and Michael Courtney, "Quantifying Momentum Transfer Due to Blast Waves from Oxy-Acetylene Driven Shock Tubes," DTIC, 2012.

Elizabeth Keenan, Sarah Warner, Ashley Crowe, and Michael Courtney, "Length, Weight, and Yield in Channel Catfish, Lake Diane, MI," Ecology, Nature Precedings, February 2011.

David Parker, Thomas Avers, and Michael Courtney, "Weight, Length, and Growth in Cutbow Trout (Oncorhynchus mykiss x clarkii)," Ecology, Nature Precedings. September 2011.

Scott Williams and Michael Courtney, "Why Cheating is Wrong," Physics Education, Cornell University Library, February 2011.